\documentclass[xcolor=pdftex,usenames,dvipsnames,table]{PoS}

\usepackage{slashed}
\usepackage{graphicx}
\usepackage{amssymb,amsmath}
\usepackage{pifont}
\usepackage{multirow,lscape}
\usepackage{array}
\usepackage{subfigure}
\usepackage{float}
\usepackage{verbatim}
\usepackage{bm}
\usepackage{comment}
\usepackage[T1]{fontenc}

\graphicspath{{./figs/}}

\providecommand{\wbar}[1]{\overline#1}
\newcolumntype{C}[1]{>{\centering\arraybackslash}p{#1}}
\providecommand{\order}[1]{\mathrm{O}(#1)}

\definecolor{indigo}{RGB}{75,0,130}
\definecolor{darkorange}{RGB}{255,140,0}
\definecolor{darkgreen}{RGB}{0,102,68}

\title{%
	Update on Heavy-Meson Spectrum Tests of the Oktay--Kronfeld Action}

\ShortTitle{%
	Tests of the Oktay--Kronfeld Action}

\author{%
  Jon A. Bailey, \speaker{Yong-Chull Jang}$^\dagger$, Weonjong Lee$^\ddagger$ \\
  Lattice Gauge Theory Research Center, CTP, and FPRD, \\
  Department of Physics and Astronomy, 
  Seoul National University,
  Seoul, 151-747, South Korea\\
  E-mail$^\dagger$:\email{integration.field@gmail.com}\\
  E-mail$^\ddagger$:\email{wlee@snu.ac.kr}
}

\author{%
  Carleton DeTar\\
  Department of Physics and Astronomy, University of Utah,
  Salt Lake City, UT  84112, USA\\
  E-mail: \email{detar@physics.utah.edu}
}

\author{%
  Andreas S. Kronfeld\\
  Theoretical Physics Department, Fermilab, Batavia, IL  60510, USA\\
  Institute for Advanced Study, Technische Universit\"at M\"unchen,
  85748 Garching, Germany\\
  E-mail: \email{ask@fnal.gov}
}

\author{%
  Mehmet B. Oktay\\
  Department of Physics and Astronomy, University of Iowa, Iowa City, IA 52242, USA 
}

\author{Fermilab Lattice, MILC, and SWME Collaborations}

\abstract{We present updated results of a numerical improvement test with heavy-meson spectrum for the
Oktay--Kronfeld (OK) action.
The OK action is an extension of the Fermilab improvement program for massive Wilson fermions including all
dimension-six and some dimension-seven bilinear terms.
Improvement terms are truncated by HQET power counting at $\order{\Lambda^3/m_Q^3}$ for heavy-light systems,
and by NRQCD power counting at $\order{v^6}$ for quarkonium.
They suffice for tree-level matching to QCD to the given order in the power-counting schemes.
To assess the improvement, we generate new data with the OK and Fermilab action that covers both charm and
bottom quark mass regions on a MILC coarse $(a \approx 0.12~\text{fm})$ $2+1$ flavor, asqtad-staggered
ensemble.
We update the analyses of the inconsistency quantity and the hyperfine splittings for the rest and kinetic
masses.
With one exception, the results clearly show that the OK action significantly reduces heavy-quark
discretization effects in the meson spectrum.
The exception is the hyperfine splitting of the heavy-light system near the $B_s$ meson mass, where statistics are
too low to draw a firm conclusion, despite promising results.}

\FullConference{The 33rd International Symposium on Lattice Field Theory\\
                 14 -18 July  2015\\
                 Kobe International Conference Center, Kobe, Japan}

\begin{document}

\section{Introduction}

Nowadays, lattice QCD achieves very high precision for calculations of light-quark
processes~\cite{Aoki2013:hep-lat.1310.8555}.
However, simulating heavy quarks in lattice QCD is still a challenge \cite{El-Khadra:Lat13.PoS001,
Bouchard:Lat14.PoS002}.
A major challenge in increasing the precision of lattice QCD calculations of heavy-quark, $c$ and $b$,
quantities is controlling heavy-quark discretization errors.
Because the heavy-quark masses and the accessible ultraviolet cutoff $a^{-1}$ are comparable, special care
is needed to handle heavy-quark discretization effects.

The Fermilab method was introduced to address these issues~\cite{EKM}.
The Oktay-Kronfeld (OK) action~\cite{Oktay2008:PhysRevD.78.014504} is an improvement (in the Symanzik sense)
of the this approach that incorporates the dimension-six and -seven bilinear operators needed for tree-level
matching to QCD.
It explicitly treats corrections through $\order{\lambda^3}$, where $\lambda \sim \Lambda_\text{QCD}/m_Q$ or
$\Lambda_\text{QCD}a$, in HQET power counting for heavy-light mesons, and $\order{v^6}$, where $v$ is the
relative quark-antiquark velocity, in NRQCD power counting for quarkonium.
For a small mass $m_Qa \ll 1$, the improvement is equivalent to $\order{a^2}$ with some $\order{a^3}$ terms
with Symanzik power counting~\cite{Symanzik:NuclPhysB.226.187}.
Based on semiquantitative arguments, it is expected that the bottom and charm quark discretization errors
could be reduced below the current $1\%$ level with the OK action~\cite{Oktay2008:PhysRevD.78.014504}.
We aim to test the improvement quantitatively.

For the heavy-light and quarkonium spectra of $c$- and $b$-mesons, we present results for the inconsistency
quantity \cite{Collins,Kronfeld} and hyperfine splittings, which test how well the Fermilab and OK actions
reduce heavy-quark discretization errors in practice.
For the work reported here, we generate data using the tadpole-improved Fermilab and OK actions for a range
of heavy-quark masses encompassing charm and bottom.
We extend our preliminary analysis on a MILC asqtad-staggered $N_f=2+1$ coarse ensemble with
$a\approx0.12~$fm \cite{Bailey2014:LAT2014.097}.
Near both charm and bottom masses, we generate data with four (two) different values of the hopping
parameter for the OK (Fermilab) action, for comparison.
We use an optimized conjugate gradient (CG) inverter \cite{JANG:LAT2013}.
Tadpole improvement of all terms is fully implemented in this program, completing early
work~\cite{MBO:LAT2010}.

\section{Meson Correlator}
\label{sec:corr}

We use the MILC asqtad-staggered $N_f=2+1$ gauge ensemble which has dimensions $N_L^3 \times N_T = 20^3
\times 64$, $\beta=6.79$, tree-level tadpole factor $u_0=0.8688$, and lattice spacing $a \approx 0.12~$fm
\cite{Bazavov:RevModPhys.82.1349}.
The asqtad-staggered action \cite{Bazavov:RevModPhys.82.1349} is used for the light degenerate sea quarks
with mass $am_l=0.02$ and strange sea quark with mass $am_s=0.05$.
For the tests reported here, we use $N_\text{cfg}=500$ configurations of the approximately 2000
configurations available.
For each configuration, we use six sources $h(\bm{r}_i,t_i)$ for calculating valence quark propagators.
The spatial source coordinates $\bm{r}_i$ are randomly chosen within the spatial cube.
The source times $t_i$ are evenly spaced along the lattice with a randomized offset $t_0 \in [0,20)$ for
each configuration.

%
%

We compute two-point correlators as described in Ref.~\cite{Bailey2014:LAT2014.097} at 10 meson momenta,
$\bm{p} = 2\pi\bm{n}/N_La$, with $\bm{n} = (0,0,0)$, $(1,0,0)$, $(1,1,0)$, $(1,1,1)$, $(2,0,0)$, $(2,1,0)$,
$(2,1,1)$, $(2,2,0)$, $(2,2,1)$, $(3,0,0)$---including all permutations of the components.

The definition of the hopping parameter for the OK action is given in Eq.~(2.1) of Ref.~\cite{MBO:LAT2010}. For the Fermilab action, the Eq.~(2.2) of Ref.~\cite{EKM} defines the hopping parameter.
The hopping parameter values used to simulate $c$ and $b$ quarks are given in Table~\ref{tab:kappa}.
%
%
We fix the valence light quark mass for the heavy-light meson correlators to the strange sea quark mass
$am_s$.
Hence, we refer to the heavy-light mesons as ``$B_s$'' or ``$D_s$,'' depending on the hopping parameter.
In anticipation of tuning runs for the OK action, we generate $B_s$ and $D_s$ correlators by using the OK
action with four different values for the hopping parameter $\kappa_\text{OK}$.
For purposes of comparison, we simulate with the Fermilab action with two values for the hopping parameter
$\kappa_\text{FL}$ yielding quark masses in the same ranges.
\begin{table}[tb!]
    \caption{Hopping parameter $\kappa_\text{OK}$ for the OK action and $\kappa_\text{FL}$ for the Fermilab 
        action.
        The $\kappa_\text{FL}$ are vertically aligned to the $\kappa_\text{OK}$, which yield the closest 
        heavy-light meson kinetic mass $M_2$.
        See also Fig.~\protect\ref{fig:iparam}.}
  \label{tab:kappa}
  \centering
  \renewcommand{\arraystretch}{0.9}
    \begin{tabular}{l | c c c c | c c c c}
      \hline\hline
      \multicolumn{1}{c|}{$Q$} &
      \multicolumn{4}{c|}{$b$} & 
      \multicolumn{4}{c}{$c$} 
      \\ \hline
      $\kappa_\text{OK}$ & 0.039 & 0.040 & 0.041 & 0.042 & 0.0468 & 0.048 & 0.049 & 0.050 
      \\
            $\kappa_\text{FL}$ & & & 0.083 & 0.091 & & 0.121 & 0.127 &
      \\ \hline\hline
    \end{tabular}
\end{table}


The ground state energies $E$ are extracted from correlator fits to the function
\begin{align}
  \label{eq:corrfitfn}
  f(t) 
  &= A e^{-E t} \big\{1-(-1)^t r e^{-\Delta E t}\big\}
  + A e^{-E (T-t)} \big\{1-(-1)^t r e^{-\Delta E (T-t)}\big\} ,
\end{align}
where $A$ is the ground state amplitude.
We also incorporate the staggered parity partner state with amplitude $A^p$ and energy $E^p$ for the ground
state into the fit function.
In practice, we take an amplitude ratio $r=A^p/A$ and energy difference $\Delta E=E^p-E$ as fit parameters
instead of $A^p$ and $E^p$.
We set the Bayesian prior $\Delta E=0.2(5)$.
The parity partner is not involved in the fit for quarkonium, because both heavy quarks are described by
either the OK or Fermilab action.
Hence, we used a simpler fit function for quarkonium with $r \equiv 0$ in Eq.~\eqref{eq:corrfitfn}.

We perform correlated fits.
The inverse covariance matrix is estimated with singular value decomposition (SVD).
Before performing SVD, the covariance matrix is normalized by the maximum component on the diagonal, so
that the largest eigenvalue $\lambda_\text{max}$ is of $\order{1}$.
Then singular values $\lambda_i$ below the numerical tolerance $\lambda_i/\lambda_\text{max}\leq10^{-15}$
are removed.
One or two singular values are removed from the data with $b$-quark hopping parameter values given in
Table~\ref{tab:kappa}.

To increase statistics, each correlator is averaged over positive and negative time separations. 
Then, we take the fit interval $[t_\text{min},t_\text{max}]$, where $0\leq t_\text{min}<t_\text{max}<T/2$,
equal to $[10,19]$ for the heavy-light systems and $[15,20]$ for the quarkonia.
We fix these intervals for fits to all correlators, independent of the hopping parameter $\kappa$, momentum
$\bm{p}$, and action.
The $t_\text{max}$ are chosen by requiring that the noise-to-signal ratio in the two-point correlator be less than about $20\%$ for all
momenta, which, in practice, is set at the larger momenta, where the correlator is noisier.
The $t_\text{min}$ are chosen by observing the effective mass calculated from the definition
$m_\text{eff}(t) = \left[ \ln \left\{{C^M(t)}/{C^M(t+2)}\right\}\right]/2$, as well as comparing the fit
results $E$ with $m_\text{eff}(t)$.
To estimate the statistical errors, we use a single-elimination jackknife.
%

\section{Meson Masses}
\label{sec:disp}

We fit the ground state energy $E$ in Eq.~\eqref{eq:corrfitfn} for each momentum $\bm{p}$ to the
non-relativistic dispersion relation $E(\bm{p})$, including terms up to $\order{(a\bm{p})^6}$,
\begin{align}
  \label{eq:disp}
  E &= M_1 + \frac{\bm{p}^2}{2M_2} - \frac{(\bm{p}^2)^2}{8M_4^3}
      + E_4^\prime + E_6 + E_6^\prime ,\\
  \label{eq:disp-hoc}
  E_{4}^\prime &= - \frac{a^3 W_4}{6} \sum_i p_i^4 ,\quad
  E_{6} = \frac{(\bm{p}^2)^3}{16M_6^5} ,\quad
  E_{6}^\prime = - \frac{a^5 W_6}{3} \sum_i p_i^6
         + \frac{a^5 W_6^\prime}{2} \bm{p}^2 \sum_i p_i^4 ,
\end{align}
to obtain the rest mass $M_1$ and kinetic mass $M_2$ of each meson.
The mismatch between the rest and kinetic meson masses can be exploited to test the improvement of
nonrelativistically interpreted actions.
The $M_{4,6}$ are generalized masses.
The $O(3)$ rotation symmetry breaking terms are $E_4^\prime$ and $E_6^\prime$.
In the continuum limit, the coefficients $W_4, W_6^{(\prime)} \to 0$ and $M_{1,4,6} \to M_2$.
The fit parameters $M_1$, $M_2^{-1}$, $M_4^{-3}$, $M_6^{-5}$, $W_4$, $W_6$, and $W_6^\prime$ are obtained by
with a linear fit.
We use the full covariance matrix among all momenta.

%
%
We investigate variations by excluding some or all the higher-order correction terms $E_4^\prime$ and $E_6^{(\prime)}$.
We find that the $E_4^\prime$ term is necessary, and $E_6^{(\prime)}$ terms reduce the $\chi^2$. 
We also investigate fits by dropping high-momentum data.
For the improvement tests, we select the results from the dispersion fits using only the lowest eight momenta without Bayesian constraints on the fit parameters.
The rest mass $M_1$ and the kinetic mass $M_2$ from the chosen fits are statistically consistent with other fits that we performed: the dispersion fits with Bayesian constraints on the fit parameters $W_4^\prime$ and $W_6^{(\prime)}$, and the dispersion fits with meson spectra which are obtained from uncorrelated fits for the two-point correlators.


\section{Inconsistency Parameter}
\label{sec:icp}

To assess the improvement, we use the inconsistency quantity~\cite{Collins, Kronfeld},
\begin{align}
  I &\equiv \frac{2\delta M_{\wbar{Q}q} - ({\delta}M_{\wbar{Q}Q} + {\delta}M_{\wbar{q}q})}{2M_{2\wbar{Q}q}}
  = \frac{2{\delta}B_{\wbar{Q}q} - ({\delta}B_{\wbar{Q}Q} + {\delta}B_{\wbar{q}q})}{2M_{2\wbar{Q}q}},
  \label{eq:iparam}
\end{align}
where the mass differences $\delta M_{X} \equiv M_{2X} - M_{1X}, (X=\wbar{Q}q, \wbar{Q}Q)$, are obtained
from the dispersion relation fits.
Because light quarks always have $ma\ll1$, the $\order{(ma)^2}$ distinction between rest and kinetic masses
is negligible.
We therefore omit $\delta M_{\bar{q}q} \equiv M_{2\wbar{q}q} - M_{1\wbar{q}q}$ (or $\delta B_{\bar{q}q}$).

The meson masses $M_{1,2}$ can be written as a sum of the perturbative quark masses $m_{1,2}$ and the binding energies $B_{1,2}$.
For a heavy-light system, the relation reads
\begin{align}
  \label{eq:M1M2}
  M_{1\wbar{Q}q} = m_{1\wbar{Q}} + m_{1q} + B_{1\wbar{Q}q} ,\quad
  M_{2\wbar{Q}q} = m_{2\wbar{Q}} + m_{2q} + B_{2\wbar{Q}q} ,
\end{align}
and similarly for a (light) quarkonium.
These formulas define $B_1$ and $B_2$.
Substituting them into the definition of $I$ in Eq.~\eqref{eq:iparam}, the quark masses cancel out, and we obtain the relation among binding energy differences $\delta B = B_2 - B_1$ in Eq.~\eqref{eq:iparam}.

In a relativistically invariant theory, the binding energies $B_1$ and $B_2$ are equal.
The ``inconsistency'' $I$ isolates the binding-energy difference $\delta B = B_2 - B_1 \neq 0$.
At the leading order, $\order{\bm{p}^2}$, it is due to discretization errors from the higher-dimension
operators in the action of $\order{(a\bm{p})^4}$, or $\order{v^4}$ in NRQCD power counting, which enter
$B_2$ \cite{Kronfeld}.
This leading-order inconsistency vanishes at tree-level for the OK action, but not for the
Fermilab action~\cite{Kronfeld,Bernard,Bailey2014:LAT2014.097}.
Hence, by construction, the inconsistency quantity $I$ is good for probing how well the OK action removes 
these discretization errors in the meson spectra.

\begin{figure}[tb!]
\vspace{-5mm}
  \centering
  \subfigure[]{
    \label{fig:I-PS-full}
    \includegraphics[width=0.45\textwidth]{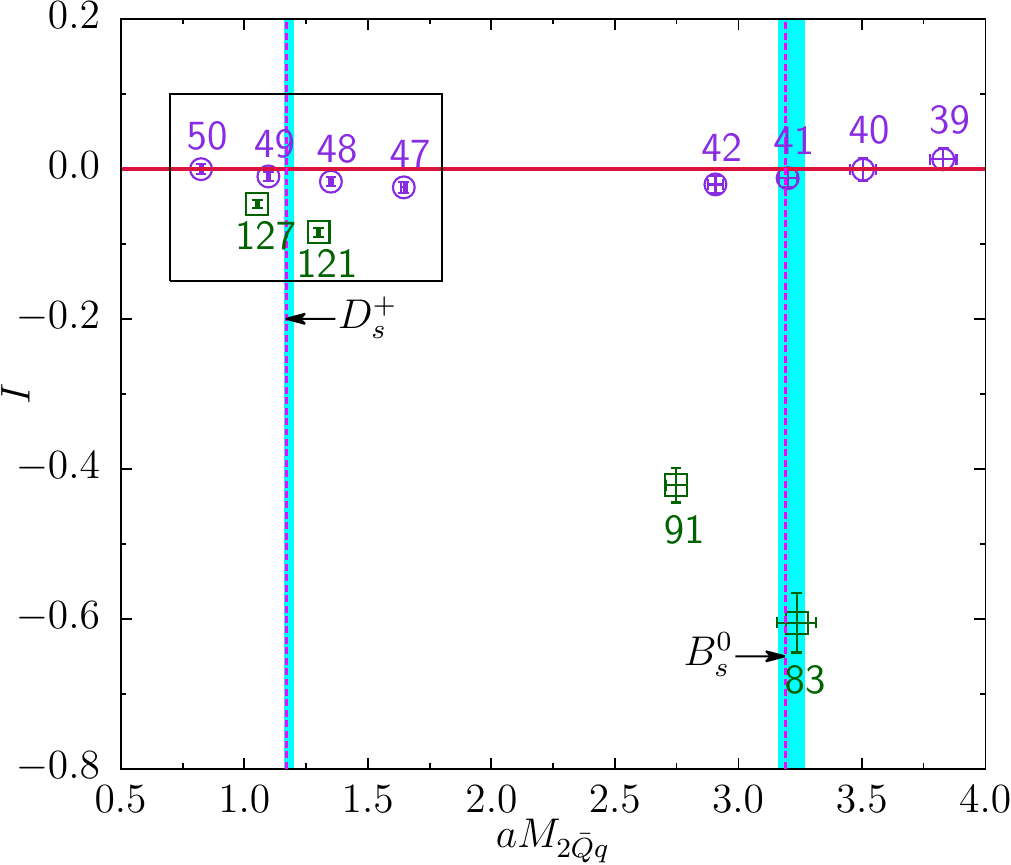}}
  \hspace{4mm}
  \subfigure[]{
    \label{fig:I-PS-c}
    \includegraphics[width=0.44\textwidth]{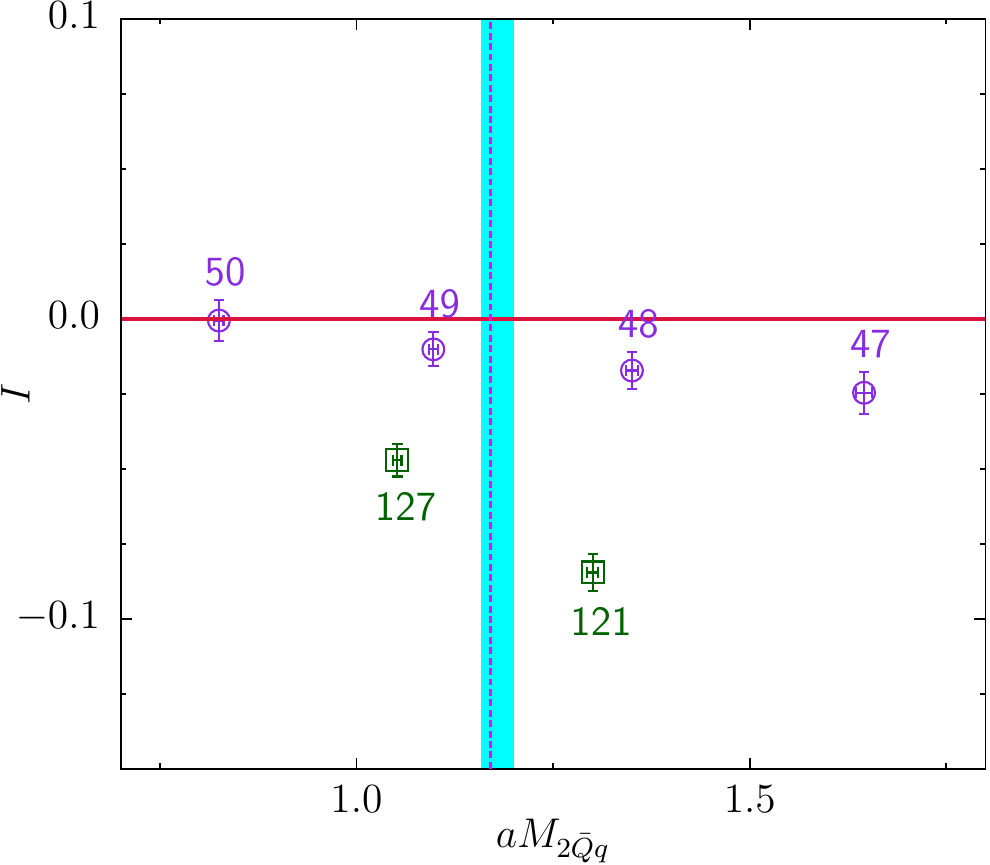}}
  \caption{Inconsistency $I$ for pseudoscalar mesons.
    A magnified view of the boxed region in Fig.~\protect\ref{fig:I-PS-full} is given in Fig.~\protect\ref{fig:I-PS-c}.
      Data labels denote $\kappa \times 10^3$ values.
      The square (green) represents the Fermilab action data, and the circle (purple) represents OK action 
      data.
      For tuning purposes, we also indicate the physical $B_s$ and $D_s$ masses with vertical lines and 
      error bands \cite{agashe2014:chinphysc.38.090001}; the errors are dominated by the error of the 
      lattice spacing.
      $I$ almost vanishes for the OK action, but for the Fermilab action it does not.
      This behavior suggests that the OK action is significantly closer to the continuum limit, $I=0$, 
      which is represented by the horizontal line (red).
      The errors are from the jackknife.}
  \label{fig:iparam}
\end{figure}

The results for the inconsistency $I$ from the pseudoscalar meson spectra are shown in Fig.~\ref{fig:iparam}.
We find that $I$ is close to the continuum limit, $I=0$, for the OK action even in the bottom mass region,
whereas the Fermilab action produces a very large deviation, $I \approx -0.6$.
The small $I$ shown in Fig.~\ref{fig:iparam} for the OK action results mainly from the higher order kinetic
operators of $\order{v^4}$ in NRQCD, or $\order{\lambda^2}$ in HQET, power
counting~\cite{Oktay2008:PhysRevD.78.014504}.
These terms suffice to tune the quark dispersion relation to $\order{(a\bm{p}^4)}$.
This outcome provides good numerical evidence that the improvement expected with the OK action is realized 
in practice.

\section{Hyperfine Splittings}
\label{sec:hfs}

The hyperfine splitting $\Delta$ is defined to be the difference in the masses of the vector ($M^\ast$) and
pseudoscalar ($M$) mesons:
\begin{equation}
  \Delta_1 = M_1^{\ast} - M_1 ,
  \qquad
  \Delta_2 = M_2^{\ast} - M_2 .
\end{equation}
Spin-independent contributions to the binding energies cancel in the difference of hyperfine splittings
$\Delta_2 - \Delta_1 = \delta{B^{\ast}} - \delta{B}$~\cite{Bailey2014:LAT2014.097}.
Comparing to the continuum limit, $\Delta_2 = \Delta_1$, diagnoses the improvement of the spin-dependent
terms in the OK action~\cite{Oktay2008:PhysRevD.78.014504} of $\order{v^6}$ in NRQCD, or $\order{\lambda^3}$
in HQET power counting.

\begin{figure}[tb!]
\vspace{-5mm}
  \centering
  \subfigure[Quarkonium]{
    \label{fig:hfs-hh}
    \includegraphics[width=.45\textwidth]{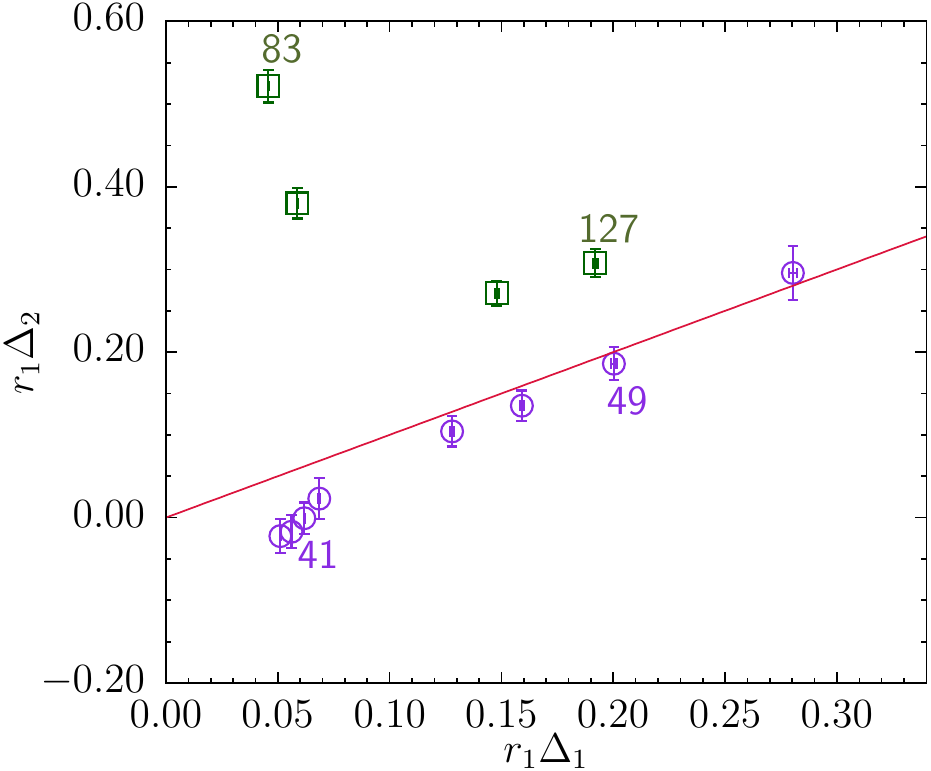}}
  \hfill
  \subfigure[Heavy-light meson]{
    \label{fig:hfs-hl}
    \includegraphics[width=.45\textwidth]{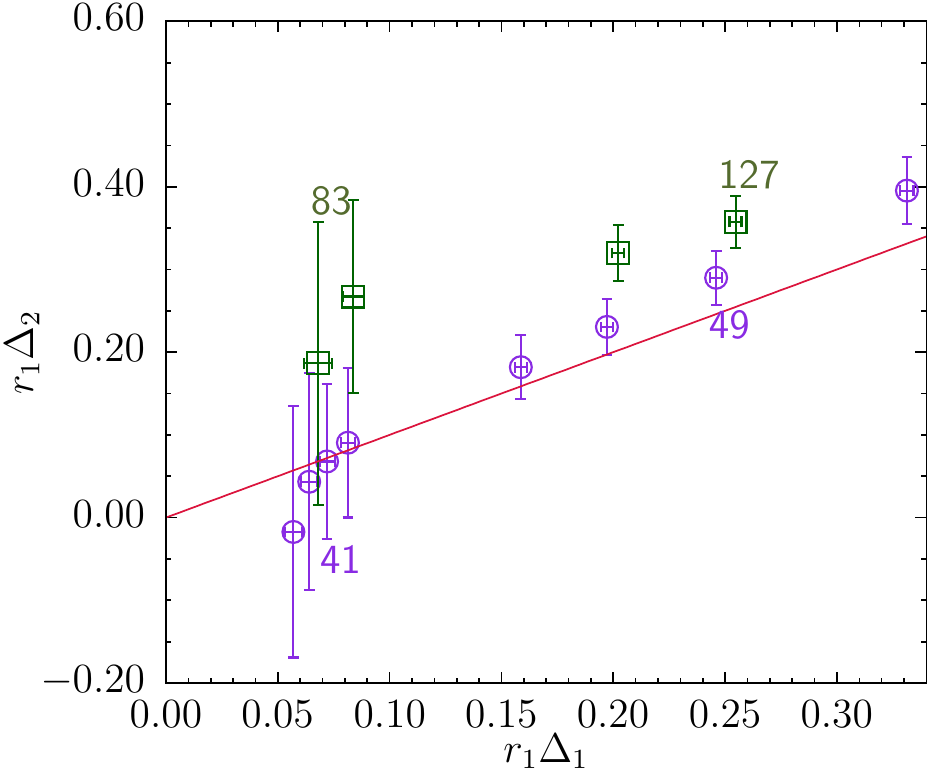}}
    \caption{Hyperfine splitting $\Delta_2$ obtained from the kinetic masses vs.\ $\Delta_1$, that obtained 
        from the rest masses.
        The square (green) represents the Fermilab action data, and the circle (purple) represents OK 
        action data.
        The labels are $\kappa \times 10^3$, corresponding to kinetic masses close to the physical $B_s$ 
        (83, 41) and $D_s$ (49, 127) masses, as shown in Fig.~\protect\ref{fig:iparam}.
        The continuum limit is represented by the line (red) $\Delta_2=\Delta_1$.
        Errors are estimated with the jackknife method.}
    \label{fig:hfs}
\end{figure}

As one can see in Fig.~\ref{fig:hfs-hh}, the OK action shows clear improvement for quarkonium.
The data points from the OK action lie much closer to the continuum limit $\Delta_2 = \Delta_1$ (the red
line) for all simulated values of $\kappa_\text{OK}$.
In addition, the deviation is smaller for the charmonium region, near $\kappa_\text{OK} = 0.049$ and
$\kappa_\text{FL} = 0.127$, than for the bottomonium region, near $\kappa_\text{OK} = 0.041$ and
$\kappa_\text{FL} = 0.083$.
The heavy-light results in Fig.~\ref{fig:hfs-hl} also show clear improvement in the region near the $D_s$
mass.
The results with the OK action remain consistent with the continuum expectation throughout the $B_s$ mass
region, but the improvement is not significant for $\kappa_\text{OK} \leq 0.041$, because the statistical
errors are large.
Even here, however, the results are suggestive of improvement.

For both quarkonia and heavy-light mesons, the hyperfine splitting of the kinetic mass ($\Delta_2$) has a
larger error than that of the rest mass ($\Delta_1$), mainly because the kinetic mass requires correlators
with $\bm{p}\neq\bm{0}$, which are noisier than those with $\bm{p}=\bm{0}$.
The statistical errors shown in Fig.~\ref{fig:hfs} are comparable for the OK and Fermilab action, except for
$\Delta_2$ for heavy-light mesons with $\kappa_\text{OK}=0.041, 0.042$, which are interestingly 30--50\%
smaller than those with $\kappa_\text{FL}=0.083, 0.091$.

\section{Conclusion}

The inconsistency quantity shows that the OK action improves the $\order{\bm{p}^2}$ effects in the binding
energy, because it improves the $\order{\bm{p}^4}$ part of the effective Lagrangian, in
practice as well as in theory.
The hyperfine splittings clearly show the improvement from the higher-dimension chromomagnetic interaction
terms, except in the $B_s$ mass region, where statistics are at present insufficient to reach any definite
conclusion.


\section{Acknowledgments}

J.A.B.\ is supported by the Basic Science Research Program of the National Research Foundation of Korea
(NRF) funded by the Ministry of Education (2015024974).
C.D.\ is supported in part by the U.S.\ Department of Energy under grant No.\ DE-FC02-12ER-41879 and the
U.S.\ National Science Foundation under grant PHY10-034278.
A.S.K.\ is supported in part by the German Excellence Initiative and the European Union Seventh Framework
Programme under grant agreement No.~291763 as well as the European Union's Marie Curie COFUND program.
Fermilab is operated by Fermi Research Alliance, LLC, under Contract No.\ DE-AC02-07CH11359 with the U.S.\
Department of Energy.
The research of W.L.\ is supported by the Creative Research Initiatives Program (No.~2015001776) of the NRF
grant funded by the Korean government (MEST).
W.~L.\ would like to acknowledge the support from KISTI supercomputing center through the strategic support
program for the supercomputing application research [No.~KSC-2014-G3-002].
The computations were carried out in part on the DAVID GPU clusters at Seoul National University.

\bibliography{refs}

\providecommand{\href}[2]{#2}\begingroup\raggedright\begin{thebibliography}{10}

\bibitem{Aoki2013:hep-lat.1310.8555}
S.~Aoki, Y.~Aoki, C.~Bernard, T.~Blum, G.~Colangelo, {\em et~al.},
  \href{http://xxx.lanl.gov/abs/1310.8555}{{\tt 1310.8555}}.

\bibitem{El-Khadra:Lat13.PoS001}
A.~X. El-Khadra, {\em PoS} {\bf LATTICE2013} (2014) 001,
  [\href{http://xxx.lanl.gov/abs/1403.5252}{{\tt 1403.5252}}].

\bibitem{Bouchard:Lat14.PoS002}
C.~M. Bouchard, {\em PoS} {\bf LATTICE2014} (2015) 002,
  [\href{http://xxx.lanl.gov/abs/1501.03204}{{\tt 1501.03204}}].

\bibitem{EKM}
A.~X. El-Khadra, A.~S. Kronfeld, and P.~B. Mackenzie, {\em Phys. Rev.} {\bf
  D55} (1997) 3933--3957, [\href{http://xxx.lanl.gov/abs/hep-lat/9604004}{{\tt
  hep-lat/9604004}}].

\bibitem{Oktay2008:PhysRevD.78.014504}
M.~B. Oktay and A.~S. Kronfeld, {\em Phys. Rev.} {\bf D78} (2008) 014504,
  [\href{http://xxx.lanl.gov/abs/0803.0523}{{\tt 0803.0523}}].

\bibitem{Symanzik:NuclPhysB.226.187}
K.~Symanzik, {\em Nucl.Phys.} {\bf B226} (1983) 187.

\bibitem{Collins}
S.~Collins, R.~Edwards, U.~M. Heller, and J.~Sloan, {\em Nucl. Phys. B Proc.
  Suppl.} {\bf 47} (1996) 455--458,
  [\href{http://xxx.lanl.gov/abs/hep-lat/9512026}{{\tt hep-lat/9512026}}].

\bibitem{Kronfeld}
A.~S. Kronfeld, {\em Nucl. Phys. B Proc. Suppl.} {\bf 53} (1997) 401--404,
  [\href{http://xxx.lanl.gov/abs/hep-lat/9608139}{{\tt hep-lat/9608139}}].

\bibitem{Bailey2014:LAT2014.097}
J.~A. Bailey, Y.-C. Jang, W.~Lee, C.~DeTar, A.~S. Kronfeld, {\em et~al.}, {\em
  PoS} {\bf LATTICE2014} (2014) 097,
  [\href{http://xxx.lanl.gov/abs/1411.1823}{{\tt 1411.1823}}].

\bibitem{JANG:LAT2013}
Y.-C. Jang {\em et~al.}, {SWME, MILC, Fermilab Lattice}, {\em PoS} {\bf
  LATTICE2013} (2014) 030, [\href{http://xxx.lanl.gov/abs/1311.5029}{{\tt
  1311.5029}}].

\bibitem{MBO:LAT2010}
C.~DeTar, A.~Kronfeld, and M.~Oktay, {\em PoS} {\bf LATTICE2010} (2010) 234,
  [\href{http://xxx.lanl.gov/abs/1011.5189}{{\tt 1011.5189}}].

\bibitem{Bazavov:RevModPhys.82.1349}
A.~Bazavov {\em et~al.}, {\em Rev. Mod. Phys.} {\bf 82} (2010) 1349--1417,
  [\href{http://xxx.lanl.gov/abs/0903.3598}{{\tt 0903.3598}}].

\bibitem{Bernard}
C.~Bernard {\em et~al.}, {Fermilab Lattice Collaboration, MILC Collaboration},
  {\em Phys. Rev.} {\bf D83} (2011) 034503,
  [\href{http://xxx.lanl.gov/abs/1003.1937}{{\tt 1003.1937}}].

\bibitem{agashe2014:chinphysc.38.090001}
K.~A. Olive {\em et~al.}, {Particle Data Group}, {\em Chin. Phys.} {\bf C38}
  (2014) 090001.

\end{thebibliography}\endgroup

\end{document}